\documentclass[aps,prc,a4paper,nofootinbib,preprintnumbers,superscriptaddress,showpacs,showkeys,twocolumn]{revtex4}

\usepackage{amsmath}
\usepackage{amssymb}
\usepackage{bm}
\usepackage{graphicx}
\usepackage{color}
\usepackage[english]{babel}

\definecolor{darkpastelgreen}{rgb}{0.01, 0.75, 0.24}

\newcommand{\corr}[1]{\textcolor{black}{#1}}

\newcommand{\be}{\begin{equation}}
\newcommand{\ee}{\end{equation}}
\newcommand{\ba}{\begin{eqnarray}}
\newcommand{\ea}{\end{eqnarray}}
\newcommand{\eq}[1]{(\ref{#1})}

\newcommand{\bs}{\boldsymbol}
\newcommand{\avr}[1]{\left\langle #1\right\rangle}
\newcommand{\Z}{{\mathbb Z}}

\def\bbbone{{\mathchoice {\rm 1\mskip-4mu l} {\rm 1\mskip-4mu l} {\rm 1\mskip-4.5mu l} {\rm 1\mskip-5mu l}}}

\begin{document}

\title{Topological susceptibility, divergent chiral density and phase diagram of chirally imbalanced QCD medium at finite temperature}

\author{Marco Ruggieri}\email[]{ruggieri@lzu.edu.cn}
\affiliation{School of Nuclear Science and Technology, Lanzhou University, 222 South Tianshui Road, Lanzhou 730000, China}
\author{Maxim N. Chernodub}\email[]{maxim.chernodub@idpoisson.fr}
\affiliation{Institut Denis Poisson UMR 7013, Universit\'e de Tours, 37200 France}
\affiliation{Pacific Quantum Center, Far Eastern Federal University, Sukhanova 8, Vladivostok, 690950, Russia}
\author{Zhen-Yan Lu}\email[]{luzhenyan@itp.ac.cn}
\affiliation{School of Physics and Electronic Science, Hunan University of Science and Technology, Xiangtan 411201, China}

\begin{abstract}
We show that the nonlocal two-flavor Nambu--Jona-Lasinio model predicts the enhancement of both chiral and axial symmetry breaking as the chiral imbalance of hot QCD matter, regulated by a chiral chemical potential $\mu_5$, increases. The two crossovers
are reasonably close to each other in the range of $\mu_5$ examined here and the pseudocritical temperatures rise with $\mu_5$. 
The curvatures of the chiral and axial crossovers for the chiral quark chemical potential approximately coincide and give $\kappa_5 \simeq - 0.011$. We point out that the presence of $\mu_5$ in thermodynamic equilibrium is inconsistent with the fact that the 
chiral charge is not a Noether-conserved quantity for massive fermions. 
The chiral chemical potential should not, therefore, be considered as a true chemical potential that sets a thermodynamically stable environment in the massive theory, but rather than as a new coupling that may require a renormalization in the ultraviolet domain. 
The divergence of an unrenormalized chiral density, \corr{coming from zero-point fermionic fluctuations,}
is a consequence of this property. We propose a solution to this problem via a renormalization procedure.
\end{abstract}

\date{April 17, 2020}



\maketitle

\section{Introduction}
 
The chirality plays an important role in the fundamental theory of strong interactions of quarks and gluons described by Quantum Chromodynamics (QCD). In QCD with $N_f$ massless quarks, the classical formulation of QCD enjoys the invariance under $SU(N_f)_{L/R}$ chiral transformations $\psi_{L/R} \to U_{L/R}\psi_{L/R}$, which transforms, separately, the $N_f$-plets of quarks with left-handed, $\psi_L$, and right-handed, $\psi_R$, chiralities. As a consequence of the Noether theorem, the chiral charge is a conserved quantity at the level of classical equations of motion.
In the quantum version of QCD, the chirality is no more a conserved number because quantum fluctuations break the chiral symmetry spontaneously. The chiral condensate, $\avr{\bar\psi\psi} \equiv \avr{\bar\psi_L\psi_R} + \avr{\bar\psi_R\psi_L}$, dynamically breaks the full chiral group down to its diagonal (vector) subgroup, $SU(N_f)_{L} \times SU(N_f)_{R} \equiv SU(N_f)_{V} \times SU(N_f)_{A}\to SU(N_f)_{V}$. Given the number of the light quarks, $N_f = 2$, the chiral symmetry breaking manifests itself in the appearance of three Goldstone bosons, the light pseudoscalar mesons, that correspond to the number of spontaneously broken generators, $2N_f - 1$, of the chiral $SU(N_f)_A$ subgroup~\cite{Shifman:1988zk}.
At a classical level, QCD also possesses the axial symmetry which reflects the invariance of the QCD Lagrangian under the axial $U(1)_A$ transformations that rotate all quark flavors by the same, chirality-sensitive phase: $\psi_{L/R} \to e^{\pm i \omega}\psi_{L/R}$. This symmetry is broken at the quantum level by the axial anomaly, which leads to nonconservation of the (otherwise, classically conserved) axial current. The chiral and axial symmetries of quarks are intrinsically related to the topological properties of the vacuum which are determined by the gluonic sector of the theory~\cite{Shifman:1988zk}. 

It is established that at zero baryon chemical potential the QCD medium experiences a smooth crossover from the hadronic, low-temperature phase characterized by a nonzero chiral condensate, to the high-temperature phase of quark-gluon plasma where the chiral condensate is 
almost vanishing. As a result, the chiral symmetry gets approximately restored with the rise of the temperature. 
The fact that this is a crossover rather than a transition is related to the nonzero quark masses that break
chiral symmetry explicitly, so the chiral condensate serves as an approximate order parameter of the chiral symmetry breaking.
There are also arguments why the axial symmetry should also be restored at high temperature~\cite{Schafer:2004gy}. The nonconservation of the axial current is related to the presence of topologically nontrivial configurations, usually associated with the instantons. 
The strength of the instanton fluctuations, quantitatively determined by the topological susceptibility, depends on the environment: 
at high temperatures instantons-based interactions among quarks are suppressed. 
Therefore, one expects that the topological susceptibility is small at high temperatures and the axial symmetry, similarly to the chiral symmetry, gets restored at the high-temperature phase. 
So far a clear relation between chiral and axial symmetry restoration is not very clear.
Although the topological susceptibility is not an order parameter, 
it seems natural to consider this to be a relevant quantity for quantifying the axial symmetry breaking,
and to define the crossover for axial symmetry restoration as the temperature range in which the topological
susceptibility has the largest change with temperature. This is the strategy that we adopt in the present article,
in which we report on the study of the restoration of chiral and axial symmetry,
studying the latter by means of the topological susceptibility.

In this article we model a thermal QCD medium at finite chiral chemical potential, $\mu_5$,
the latter being conjugated to the asymmetry between left- and right-handed particles density, $n_5$. 
The chiral sector of QCD affects also various transport properties of the system due to the presence of the axial anomaly. The most famous example of such phenomena is the Chiral Magnetic Effect (CME) which generates an electric current~\cite{Kharzeev:2007jp,Fukushima:2008xe,Fukushima:2010vw,Fukushima:2009ft,Fukushima:2010fe,Fukushima:2010zza}:
$\left\langle\bm{J} \right\rangle = (e^2/2\pi^2) \mu_5\bm{B},$ in the presence of an external magnetic field $\bm{B}$ in a system of massless fermions possessing a nonzero chiral density $n_5 \neq 0$. 
The chiral density corresponds to the difference in densities of quarks with right- and left-handed chiralities, $n_5 = n_R - n_L$, encoded in the difference between their chemical potentials, $\mu_5 = \mu_L - \mu_R$. The chiral chemical potential $\mu_5$ is thermodynamically conjugated to the chiral density $n_5$, so that in the thermal equilibrium, the finite density of massless fermions is set by the help of the matter source term $\delta L = \mu_5 n_5$. The chiral density is the temporal component of the chiral (axial) 4-current:
\begin{equation}
j^\mu_5 \equiv (n_5, {\bs j}_5) = {\bar \psi} \gamma^\mu \gamma^5 \psi.
\label{eq:j:5}
\end{equation}
The CME is suggested to play an essential role in a wide number of physical systems ranging from astrophysical systems and quark-gluon plasmas to chiral materials~\cite{Kharzeev:2013ffa}. Although the CME does not exist in a thermodynamic equilibrium~\cite{VF13,Yamamoto:2015fxa,Zubkov:2016tcp}, the electric CME current is a non-dissipative quantity even in the presence of strong interactions thanks to the topological protection~\cite{Kharzeev:2013ffa}.
In the electromagnetic sector, the nonzero chiral density ($\mu_5 \neq 0$) may be induced via the axial anomaly that creates an imbalance in densities between right- and left-handed chiral quarks in the near-equilibrium background of parallel electric $\bm E$ and magnetic $\bm B$ fields. In the context of QCD, the chiral density may also appear in the gluon sector, due to topological transitions between different vacuum states mediated by the instanton or sphaleron phase transitions. 
The chiral density depletes due to mass effects, pion and sigma exchanges at low temperature~\cite{Ruggieri:2016asg}, 
and the Compton scattering at high temperature~\cite{Manuel:2015zpa}.
Regardless of the microscopic processes involved, the use of $\mu_5$ is appropriate as long as one considers a thermodynamic system on a time scale larger than that of the equilibration time.

 In this study,
the interaction among quarks is mimicked by a nonlocal Nambu--Jona-Lasinio 
model~\cite{Nambu:1961tp,Nambu:1961fr,
Klevansky:1992qe,Hatsuda:1994pi,Buballa:2003qv,Schmidt:1994di,Bowler:1994ir,Plant:1997jr,Blaschke:2000gd,
GomezDumm:2005hy,Frasca:2011bd,Frasca:2013kka,Pagura:2016pwr,
Hell:2008cc,Hell:2009by,Langfeld:1996ac,Langfeld:1996rn}.  
The nonlocal NJL models have revealed to be appropriate for the study of the chiral medium at finite temperature,  
in particular because they predict that the critical temperature for chiral symmetry restoration, $T_c$, increases with 
$\mu_5$ \cite{Ruggieri:2016ejz,Frasca:2016rsi}, differently from what has been found within models with a local interaction 
kernel \cite{Gatto:2011wc,Fukushima:2010fe,
Chernodub:2011fr,Ruggieri:2011xc,Yu:2015hym}, 
and in agreement with lattice simulations~\cite{Braguta:2015owi,Braguta:2015zta,Astrakhantsev:2019wnp} 
and calculations based on Schwinger-Dyson equations~\cite{Xu:2015vna,Wang:2015tia}. 
We report on the calculation of $T_c$ versus $\mu_5$, confirming the previous findings that $T_c$ increases with the chiral
chemical potential at least as long as $\mu_5$ is smaller than the typical ultraviolet scale of the model 
(for large values of $\mu_5$ one should include the backreaction on the interaction kernel, but the computation of this
 is well beyond the scope of our study). 
Part of the study presented here is devoted to the divergence of $n_5$ coming from ultraviolet fermion modes:
we discuss how this divergence arises from the tail of the quark mass function, and how this divergence can be cured via
a renormalization procedure.
Then we compute the topological susceptibility as a function ot $T$ and $\mu_5$ and try to relate this to a possible crossover 
from a low temperature phase in which axial symmetry is broken by axions, to  the high 
temperature phase in which axial symmetry is partly restored; we also comment on the simultaneity of this crossover and the chiral one,
finding that the two have substantial overlap suggesting that restoration of chiral symmetry is accompanied by the restoration 
of the axial symmetry.

The structure of this paper is as follows. In Section II we present the nonlocal NJL model used in the study,
discussing also a renormalization of $n_5$. 
In Section III we present our results about chiral and axial symmetry at finite temperature and $\mu_5$.
Finally in Section IV we collect our conclusions.

\section{The model}

\subsection{Thermodynamic potential}

In this work we use a nonlocal version of the Nambu-Jona-Lasinio (NJL) model~\cite{Hell:2008cc}. As one of the main goals of this work is to evaluate the topological susceptibility, it is necessary to introduce the $\theta$-dependence of the thermodynamic potential. In the QCD Lagrangian, the topological angle $\theta$ appears as a $CP$-odd term for the gluon fields:
\begin{equation}
\delta {\mathcal L}_\theta = \theta Q, 
\label{eq:L:theta}
\end{equation}
where the topological number of the gluonic field configuration is computed via the field-strength tensor $F_{\mu\nu}^a$:
\begin{eqnarray}
Q \equiv \Delta N_{\mathrm{CS}} & \equiv & N_{\mathrm{CS}} {\Bigl |}_{t \to - \infty} - N_{\mathrm{CS}}  {\Bigl |}_{t \to + \infty} \nonumber \\
& = & \frac{g^2}{64\pi^2}\int d^4 x \, \varepsilon^{\mu\nu\rho\sigma}F^a_{\mu\nu} F^a_{\rho\sigma}.
\label{eq:CS}
\end{eqnarray}
The topological number~\eq{eq:CS} is given by the difference the Chern-Simons charge $N_{\mathrm{CS}} = N_{\mathrm{CS}}(t)$ at initial and final gluonic configurations.
  
After performing a chiral rotation of the quark fields, the topological term~\eq{eq:L:theta} disappears and the $\theta-$dependence is transmitted from the gluonic topological term~\eq{eq:L:theta} to the quark sector in the form of the chiral (axial) chemical potential $\mu_5 = \partial \theta/\partial t$.  In our work, we use the NJL model to describe the dynamics of quarks. The gluon sector will thus leave its imprint only in the phenomenological interactions between quarks and in the mentioned chiral chemical potential.

The Lagrangian density of the quark model that we use in this study is given by the sum of the four terms:
\begin{equation}
{\cal L} = {\cal L}_q + {\cal L}_m +{\cal L}_4,
\label{eq:L1}
\end{equation}
where
\begin{equation}
{\cal L}_q =\bar\psi(i\gamma^\mu\partial_\mu   +\mu_5 \gamma^0\gamma^5)\psi 
\equiv \bar\psi i\gamma^\mu\partial_\mu \psi + \mu_5 n_5,
\label{eq:Lq}
\end{equation}
denotes the free quark contribution with the chiral chemical potential $\mu_5$ and the term
\begin{equation}
{\cal L}_m = -m_0\bar\Psi\Psi,
\label{eq:L:m}
\end{equation}
introduces the current quark mass~$m_0$ via the quark field dressed by interactions:
\begin{equation}
\Psi(x) = \int~d^4y~G(x-y)\psi(y).
\label{eq:Psi}
\end{equation}
We need to introduce this special form of the mass term, Eqs.~\eq{eq:L:m} and \eq{eq:Psi}, instead of the standard term ${\cal L}_m^{(0)} = -m_0\bar\psi\psi$, because we aim to model the perturbative tail of the current quark mass computed in the perturbative QCD (pQCD) at large Euclidean momentum. The perturbative matching will appear in the Fourier transform of the form-factor $G(z)$ in Eq.~\eq{eq:Psi} at large momentum~$p$.

The last term in the quark Lagrangian~\eq{eq:L1} is a nonlocal interaction term that mimics the gluon-exchange effects:
\begin{eqnarray}
 {\cal L}_4 &=&G_1\sum_{\ell=0}^3\left[(\bar Q \tau_\ell Q)^2 + (\bar Q i\gamma^5 \tau_\ell Q)^2\right]\nonumber\\
  &&+8G_2 \left[e^{i\theta}\mathrm{det}(\bar Q_R Q_L) +e^{-i\theta}\mathrm{det} (\bar Q_L Q_R)\right].
  \label{eq:L4}
\end{eqnarray}
where $\tau_l = (\bbbone, i{\bs \tau})$ is a quaternion and the spinor $Q$ represents yet another dressed (nonlocal) quark field,
\begin{equation}
Q(x) = \int~d^4y~F(x-y)\psi(y),
\label{eq:form_factor}
\end{equation}
expressed via the formfactor $F(x-y)$ to be specified later. This formfactor is similar to the function $G(x-y)$ that appears in the other form of the dressed quark field~\eq{eq:Psi} used in the mass term of the quark Lagrangian~\eq{eq:L:m}.
This interaction has been considered in its local version in \cite{Lu:2018ukl}, see also references therein, therefore we remind
to that study for further details.

It is convenient to perform the chiral rotation of the quark fields:
\begin{equation}
\psi_R \rightarrow e^{-i\theta/4}\psi_R ,~~~\psi_L \rightarrow e^{i\theta/4}\psi_L,
\end{equation}
which removes the $\theta$-dependence in the interaction Lagrangian~\eq{eq:L4}:
\begin{eqnarray}
 {\cal L}_4 &=&G_1\sum_{\ell=0}^3\left[(\bar Q  \tau_\ell Q)^2 +
 (\bar Q  i\gamma^5 \tau_\ell Q)^2\right]\nonumber\\
  &&+8G_2 \left[ \mathrm{det}(\bar Q_R  Q_L)
  + \mathrm{det} (\bar Q_L  Q_R)\right].\label{eq:L42}
\end{eqnarray}
The $\theta$-dependence reappears in the quadratic part of the Lagrangian~\eq{eq:L1} in terms of the new fermionic fields\footnote{For the notational convenience, we use the same symbols for the old and new fields, since the old ones will not appear again.}:
\begin{eqnarray}
{\cal L}_q + {\cal L}_m &=& 
\bar\psi(i\gamma^\mu\partial_\mu  +\mu_5 \gamma^0\gamma^5)\psi \nonumber\\
&&-\bar\Psi(  m_{0+} +i m_{0-}\gamma^5  )\Psi,
\label{eq:Lq2}
\end{eqnarray}
via the rotated current masses:
\begin{eqnarray}
m_{0+} &=& m_0 \cos(\theta/2),\\
m_{0-} &=& m_0 \sin(\theta/2).
\end{eqnarray}

It is also convenient to introduce the collective fields
\begin{eqnarray}
\sigma &=& G_+\bar Q  Q,\\
\eta &=& G_-\bar Q i\gamma^5 Q;
\label{eq:eta}
\end{eqnarray}
we have put $G_\pm = G_1 \pm G_2$. Following the established procedure of bosonization at the one-loop approximation we get the thermodynamic potential,
\begin{eqnarray}
& & \Omega = \frac{\sigma^2}{G_+} +\frac{\eta^2}{G_-} \label{eq:omega_1} \\
 && \quad -N_c N_f T\sum_n\int\frac{d^3p}{(2\pi)^3}\log\beta^4\left(\omega_n^2 + E_+^2\right) \nonumber
 \left(\omega_n^2 + E_-^2\right), 
\end{eqnarray}
where $\beta=1/T$ is the inverse temperature and $\omega_n=\pi T(2n+1)$ with $n \in \Z$ are the fermionic Matsubara frequencies. In Eq.~\eq{eq:omega_1} have also defined the energy branches:
\begin{equation}
E^2_\pm(p)  = (|\bm p|\pm\mu_5)^2 + {\cal M}^2(p) + {\cal N}^2(p),
\label{eq:E2:pm}
\end{equation}
with
\begin{eqnarray}
{\cal M}(p) &=& m_{0+}{\cal R}(p)  -2{\cal C}(p)\sigma,\label{eq:mms:M}\\
{\cal N}(p) &=& m_{0-}{\cal R}(p)  -2{\cal C}(p)\eta.\label{eq:mms:N}
\end{eqnarray}
Here the function ${\cal C}(p)\equiv F^2(p)$ is determined via the Fourier transform $F(p)$ of the form factor in Eq.~\eqref{eq:form_factor}.
The function ${\cal R}(p) = G(p)^2$ gives the evolution of the (renormalized) 
current quark mass with the UV scale that is necessary to reproduce the independence of the combination $m \langle\bar qq\rangle$ from the renormalization point.

The expectation (i.e., the mean field) values of the condensates $\sigma$ and $\eta$ are determined at each temperature $T$ and chiral chemical potential $\mu_5$ by the minimization of the thermodynamic potential $\Omega$. For sake of notational convenience, we have used $\sigma$ and $\eta$ to denote the mean field values of these fields. To derive Eq.~\eqref{eq:omega_1}, we used the imaginary time formalism to deal with the finite temperature bath and employed the analytical continuation to Euclidean momentum $p_E = ({\bs p}, p_4 = - i p_0)$.

For future reference it is useful to define the quantities
\begin{eqnarray}
M(p_E) &=& m_0 {\cal R}(p_E) - 2 \sigma{\cal C}(p_E),\\
m(p_E) &=& m_0 {\cal R}(p_E),
\end{eqnarray}
which correspond to the quark mass function and the current mass at $\theta=0$, respectively.

We now specify the analytical forms of ${\cal R}(p)$ and ${\cal C}(p)$. For the latter we follow~\cite{Ruggieri:2016ejz} and  take
\begin{eqnarray}
{\cal C}(p_E) &=& \theta(\Lambda^2 - p_E^2) \nonumber\\
 &+& \theta(p_E^2 -\Lambda^2) \frac{\Lambda^2}{p_E^2}
\frac{\left(\log \Lambda^2/\Lambda^2_\mathrm{QCD}\right)^\gamma}
{\left(\log p_E^2/\Lambda^2_\mathrm{QCD} \right)^{\gamma}};\nonumber\\
&&\label{eq:CPE}
\end{eqnarray}
here $p_E$ is the Euclidean 4-momentum and $\gamma=1-d_m$ is given by the anomalous dimension of the current quark mass for a two-flavor QCD:
\begin{equation}
d_m = 12/29.
\label{eq:d:m}
\end{equation}
The second line mimics the quark mass function computed in perturbative QCD
arising from the chiral condensate \cite{Politzer:1976tv,Langfeld:1996ac,Langfeld:1996rn}, 
while $\Lambda$ is an additional parameter of the model that corresponds
to the momentum scale at which the perturbative mass is matched to the nonperturbative one.
Differently from previous works, we have also introduced an energy scale dependence of the current mass,
that mimics the running of this quantity computed in perturbative QCD \cite{Politzer:1976tv,Langfeld:1996ac,Langfeld:1996rn}:
 \begin{eqnarray}
{\cal R}(p_E) &=& \theta(\Lambda^2 - p_E^2) \nonumber\\
 &+& \theta(p_E^2 -\Lambda^2)
\frac{\left(\log \Lambda^2/\Lambda^2_\mathrm{QCD}\right)^{d_m}}
{\left(\log p_E^2/\Lambda^2_\mathrm{QCD} \right)^{d_m}}.\nonumber\\
&&\label{eq:RPE}
\end{eqnarray}
The perturbative $p_E$ tails in the functions ${\cal C}(p_E)$ and ${\cal R}(p_E)$ ensure that the divergence of the chiral condensate
with a ultraviolet (UV) cutoff $\Lambda_\mathrm{UV}\gg\Lambda$ is absorbed by that of the current mass so that the combination $m_0 \langle\bar qq\rangle$
is independent from the UV cutoff~$\Lambda_\mathrm{UV}$. Moreover,
the log-tail of the current quark mass will make the divergence of $n_5$ softer than that of a fermion
gas with a momentum independent mass, as we discuss in Section~III. 

Strictly speaking, the thermodynamic potential~\eqref{eq:omega_1} is a UV divergent quantity which has to be regularized at a proper subtraction point. In this study, we chose to subtract the potential with the vanishing condensates $\sigma=\eta=0$ at $T=\mu_5=0$ and $\theta\neq 0$, namely at the free vacuum Fermi gas contribution at a finite topological angle~$\theta$. Therefore, we work with the following form of th UV-regularized thermodynamic potential:
\begin{eqnarray}
\Omega &=& \frac{\sigma^2}{G_+} +\frac{\eta^2}{G_-}\nonumber\\
 &&-N_c N_f T\sum_n\int\frac{d^3p}{(2\pi)^3}\log\beta^4\left(\omega_n^2 + E_+^2\right)
 \left(\omega_n^2 + E_-^2\right) \nonumber\\
 && +2N_c N_f \int\frac{d^4p_E}{(2\pi)^4}\log\left(\frac{p_E^2 + m_{0+}^{2}}{\Lambda^2}\right).
 \label{eq:omega_subtr}
\end{eqnarray}
We added the factor $\Lambda$ in the denominator of the last $\log$ function for the sake of the dimensional consistency of the equation. This addition is irrelevant for the computation of any physical quantity.

\subsection{Chiral condensate}

Before proceeding with the actual calculations, we discuss subtleties of the definition of the chiral condensate. We firstly focus on the $CP$-invariant case $\mu_5=0$ then we generalize the discussion to the case of the chiral medium with $\mu_5 \neq 0$.  Also, we ignore a possible presence of the $\eta$ condensate~\eq{eq:eta} because this condensate breaks the time reversal symmetry ($T$: $t \to - t$) while the chiral chemical potential is a $T$-even quantity.

In general, we can write for each quark flavor $q$:
\begin{equation}
\langle\bar qq\rangle = -\mathrm{Tr}\left(S-S_0\right),\label{eq:cc1}
\end{equation}
where $S$ corresponds to the full quark propagator and $S_0$ denotes the propagator of quarks with only the current mass
taken into account. The subtraction in Eq.~\eqref{eq:cc1} is necessary to take into account only the contribution
to $\langle\bar qq\rangle $ that comes from the interaction and not from the current quark mass.

It is well known that the chiral condensate diverges  logarithmically with the renormalization scale in perturbative 
QCD \cite{Langfeld:1996ac,Langfeld:1996rn}: 
this behavior is respected in our model. As a matter of fact we have
\begin{eqnarray}
\langle\bar qq\rangle &=& -N_c\int\frac{d^4p_E}{(2\pi)^4}\frac{4M(p_E)}
{ p_4^2 + \bm p^2+ M(p_E)^2} \nonumber \\
&&+ N_c\int\frac{d^4p_E}{(2\pi)^4}\frac{4m(p_E)}
{p_4^2 + \bm p^2+ m(p_E)^2}. \label{eq:cc2}
\end{eqnarray}
In the UV regime each component of the Euclidean momentum $p_E$ should be taken much larger than the masses $M$ and $m$:
\begin{eqnarray}
\langle\bar qq\rangle_\mathrm{UV} &\approx &  -4\int_{\Gamma}^{\Lambda_\mathrm{UV}}\frac{d^4p_E}{(2\pi)^4}\frac{[M(p_E)-m(p_E)]}
{p_E^2},\nonumber\\
&&
\end{eqnarray}
with $\Gamma>\Lambda$. From the above equation, it is clear that the subtraction in Eq.~\eqref{eq:cc1} leaves the contribution of the interaction as the only one that is taken into account for the evaluation of the chiral condensate. Moreover, a simple power counting shows that the
integral possesses a log-divergence in the limit $\Lambda_\mathrm{UV}\gg\Gamma$ due to the perturbative tail of $M(p_E)$.

The previous discussion can be generalized to a nonzero chiral chemical potential, $\mu_5\neq 0$. Instead of Eq.~\eqref{eq:cc2} we now have
\begin{eqnarray}
\langle\bar qq\rangle &=& -N_c\int\frac{d^4p_E}{(2\pi)^4}\frac{4M(p_E)[M(p_E)^2 + p_4^2 + \bm p^2 + \mu_5^2]}
{[p_4^2 + \lambda_+^2(p_E)][p_4^2 + \lambda_-^2(p_E)]} \nonumber \\
&&+ N_c\int\frac{d^4p_E}{(2\pi)^4}\frac{4m(p_E)[m(p_E)^2 + p_4^2 + \bm p^2 + \mu_5^2]}
{[p_4^2 + \phi_+^2(p_E)][p_4^2 + \phi_-^2(p_E)]}, \nonumber \\
&&\label{eq:cc3}
\end{eqnarray}
where 
\begin{eqnarray}
\lambda_\pm^2(p_E) &=& \left(|\bm p|\pm\mu_5\right)^2 + M(p_E)^2,\label{eq:lam_def}\\
\phi_\pm^2(p_E) &=& \left(|\bm p|\pm\mu_5\right)^2 + m(p_E)^2.\label{eq:phi_def}
\end{eqnarray}
Again in the UV regime in which each component of the Euclidean momentum~$p_E$ is taken much larger than the masses $M$ and $m$. At the lowest order in $\mu_5$ we get:
\begin{eqnarray}
\langle\bar qq\rangle_\mathrm{UV} &\approx &  -4\int_{\Gamma}^{\Lambda_\mathrm{UV}}\frac{d^4p_E}{(2\pi)^4}\frac{[M(p_E)-m(p_E)]}
{p_E^4} (p_E^2 + \mu_5^2),\nonumber\\
&&
\end{eqnarray}
with the infrared cutoff $\Gamma>\Lambda$. Using the power counting, we find that the chiral chemical potential gives a $\mu_5^2$ correction to the chiral condensate. This correction, proportional to $[M(p_E)-m(p_E)]/p_E^4$, is finite in the UV regime.

We finally remark that although the chiral condensate has a UV log-divergence, the condensate $\sigma=\langle\bar Q Q\rangle$ computed by minimization of the thermodynamic potential~\eq{eq:omega_subtr}, is a finite quantity independent of the ultraviolet cutoff. Indeed, it is easy to
realize that the gap equation, $\partial\Omega/\partial\sigma=0$, gives in this case
\begin{eqnarray}
\sigma = -2 N_c N_f T G_+ \sum_n\int\frac{d^3p}{(2\pi)^3}\frac{{\cal M}(\omega_n,{\bs p}) {\cal C}(\omega_n,{\bs p})}{\omega_n^2 + E_+^2},
\end{eqnarray}
where we also used Eqs.~\eq{eq:E2:pm} and \eq{eq:mms:M}.
The loop integral on the right hand side of the above equation is finite in the nonlocal NJL model due to the
form factor~\eqref{eq:CPE} which removes high--momentum modes. Note that this factor does not appear
in the chiral condensate. As a consequence, while in the local NJL model
the quark condensate is proportional to $\sigma$, in the nonlocal model this proportionality is lost \cite{Hell:2008cc}.

\subsection{Chiral density}

\subsubsection{Thermodynamic definition}

In this subsection, we discuss the chiral density within the nonlocal NJL model.
The chiral density
\begin{equation}
n_5 = -\frac{\partial\Omega}{\partial\mu_5},
\end{equation}
is given by the variation of the thermodynamic potential $\Omega$ with respect to the chiral chemical potential~$\mu_5$. From Eq.~\eqref{eq:omega_1} we get, at the minimum of $\Omega$:
\begin{eqnarray}
&&n_5 =4 N_c N_f\mu_5{\cal H},\label{eq:n5_2}
\end{eqnarray}
with
\begin{equation}
{\cal H} = T\sum_n\int\frac{d^3p}{(2\pi)^3}
\frac{\omega_n^2-\bm p^2+M(p_E)^2+\mu_5^2}
{[\omega_n^2 + \lambda_+^2(p_E)][\omega_n^2 + \lambda_-^2(p_E)]},
\label{eq:integrand}
\end{equation}
where the functions $\lambda_\pm$ are defined in Eq.~\eqref{eq:lam_def} and the Euclidean momentum is determined at the Matsubara frequencies, 
$p_E = (\omega_n,\bm p)$.

\subsubsection{Divergence of unrenormalized chiral density}

Before presenting the results on the chiral density $n_5$ obtained within the nonlocal NJL model, we find useful to make a remark on the divergence of the density $n_5$ for a case when the quark mass is a fixed finite quantity. We limit this short discussion to the zero-temperature case, $T=0$, since the finite temperature part provides us with a finite contribution; in this case, we substitute $\omega_n\rightarrow p_4$ and take the integral over the continuous momentum $p_4$ along the full real axis.

If we set the dressed mass to zero, $M=0$, in Eq.~\eqref{eq:n5_2} then the  trivial integration over the momentum $p_4$ along the full real axis gives us the following expression for the zero-temperature chiral density:
\begin{equation}
n_5 = 2N_c N_f\int\frac{d^3p}{(2\pi)^3}\theta(\mu_5 - |\bm p|)=\frac{N_c N_f\mu_5^3}{3\pi^2}.
\label{eq:array_n5}
\end{equation}   
The above equation provides us with the standard relation between a density, $n$,
and a chemical potential,  $\mu$, of an ultrarelativistic massless fermion gas.

We now consider the effect of a momentum independent  mass, $m_B$, on the chiral density $n_5$.
From Eq.~\eqref{eq:n5_2} after integrating over the momentum $p_4$ along the real axis, we get:
\begin{equation}
n_5 = 2 N_c N_f \int\frac{d^3p}{(2\pi)^3}{\cal X}(p),\label{eq:nba2}
\end{equation}
where
\begin{eqnarray}
{\cal X}(p,\mu_5) &=& \frac{|\bm p|+\mu_5}{2\sqrt{(|\bm p|+\mu_5)^2 + m_B^2}} \nonumber \\
&&-\frac{|\bm p|-\mu_5}{2\sqrt{(|\bm p|-\mu_5)^2 + m_B^2}}.\label{eq:aray_n5_7}
\end{eqnarray}
The above equation can be interpreted as a distribution function
of a quark at $T=0$, $\mu_5\neq 0$ and $m_B\neq 0$.
For a zero mass $m_B=0$, Eq.~\eq{eq:aray_n5_7} naturally leads to Eq.~\eqref{eq:array_n5}, thus implying that a Fermi sphere is filled up to 
the Fermi momentum at $|\bm p| = \mu_5$.

The effect of the presence of the mass $m_B$ is to enlarge the chiral Fermi surface by putting particles above the Fermi momentum. 
Indeed, we may take in Eq.~\eq{eq:aray_n5_7} the large spatial momentum $|\bm p|$ limit by assuming naturally the order\footnote{Assuming this specific order for $m_B$, $\mu_5$ and $|\bm p|$ is irrelevant
in the UV limit; this choice is closer to the nonlocal NJL model because of the running
of the current quark mass in the UV.} $m_B\ll\mu_5\ll|\bm p|$, and obtain:
\begin{equation}
{\cal X}(p,\mu_5)\approx \frac{m_B^2\mu_5}{|\bm p|^3}.\label{eq:agree}
\end{equation}
Despite the distribution~\eq{eq:agree} decays as fast as $p^{-3}$, the density of the states
increases proportional to the phase-volume factor $p^2$. Therefore the net contribution to the chiral density behaves as $\int dp/p$, 
thus giving rise to a logarithmic divergence of the chiral density $n_5$ in the presence of a nonzero mass, $m_B \neq 0$.

The discussion of the example with $m_B\neq 0$ paves the way for understanding of the properties of the chiral density $n_5$ in the nonlocal NJL model.
In the latter case, the integral over the momentum $p_4$ cannot be taken explicitly to the momentum dependence of the quark mass function $M = M(p_E)$.
Nevertheless, we may figure out the UV divergence of the density $n_5$ because to this end it is enough to consider
the asymptotic behavior of the integrand in Eq.~\eqref{eq:array_n5} in the high-momentum limit $p_4, p \sim \Lambda_\mathrm{UV}$
with the large ultraviolet cutoff $\Lambda_\mathrm{UV}\gg\mu_5,M$. 

We firstly expand the integrand in Eq.~\eqref{eq:integrand} in powers of the quark mass function $M$ at the lowest nontrivial order:
\begin{eqnarray}
&&\frac{p_4^2-\bm p^2+M(p_E)^2+\mu_5^2}
{[p_4^2 + \lambda_+^2(p_E)][p_4^2 + \lambda_-^2(p_E)]}\nonumber\\
&&\approx \frac{p_4^2-\bm p^2 + \mu_5^2}
{[p_4^2 + (|\bm p|-\mu_5)^2][p_4^2 +(|\bm p| + \mu_5)^2]} \label{eq:integranda_2} \\
&& +M^2(p_E)\frac{3\bm p^4 + 2\bm p^2(p_4^2 - \mu_5^2) - (p_4^2 + \mu_5^2)^2}
{[p_4^2 + (|\bm p|-\mu_5)^2]^2[p_4^2 +(|\bm p| + \mu_5)^2]^2}.
\nonumber
\end{eqnarray}
Since we are interested in the UV limit of this integrand, we can safely assume the hierarchy $M(p_E)\ll\mu_5$ which is valid due to the diminishing perturbative tail of the quark mass function~$M(p_E)$.

The first term on the right hand side of Eq.~\eqref{eq:integranda_2} leads to a finite integral and
gives back the result~\eqref{eq:array_n5} for the massless quarks while the UV divergence of the chiral density $n_5$
comes from the integral of the second term in the right hand side of Eq.~\eqref{eq:integranda_2}.
The quark mass function~$M(p_E)$ gets contributions from both the chiral condensate and the current quark mass~\eqref{eq:mms:M}. However, for a large Euclidean momentum $p_E$, the latter factor dominates since the former is suppressed by the perturbative tail $1/p_E^2$ according to Eqs.~\eqref{eq:CPE} and~\eqref{eq:RPE}.

Therefore in the high momentum limit of Eq.~\eqref{eq:integranda_2}, we can replace the quark mass function $M(p_E)$ with the mass $m(p_E)$. 
Moreover, in the UV region $p_4,|\bm p|\gg\mu_5$ we can safely set $\mu_5=0$ in the second term in the right hand side of Eq.~\eqref{eq:integranda_2}
since higher powers of chiral chemical potential $\mu_5$ in a $\mu_5/|p_E|$ expansion would only lead to convergent integrals. Thus, we may finally rewrite Eq.~\eqref{eq:integranda_2} in the following form:
\begin{eqnarray}
&&\frac{p_4^2-\bm p^2+M(p_E)^2+\mu_5^2}
{[p_4^2 + \lambda_+^2(p_E)][p_4^2 + \lambda_-^2(p_E)]}\nonumber\\
&&\approx \frac{p_4^2-\bm p^2 + \mu_5^2}
{[p_4^2 + (|\bm p|-\mu_5)^2][p_4^2 +(|\bm p| + \mu_5)^2]} \nonumber \\
&& +m^2(p_E)\frac{3\bm p^2 - p_4^2 }
{p_E^6 } + O\left(m^2(p_E)\mu_5^2/p_E^6\right).
\label{eq:integranda_3}
\end{eqnarray}
It is now easy to recognize in the last term, proportional to the mass squared, $m^2(p_E)$, the source of the UV divergence of the chiral density~$n_5$.
This divergence would be of a log-type if the mass $m$ were a constant quantity. The actual $p_E$-dependence of the mass $m = m(p_E)$ leads to somewhat smoother divergence. Indeed, taking into account the behavior of the aforementioned term at the momentum shell $\Gamma\ll|p_E|\ll\Lambda_\mathrm{UV}$, we get, ignoring an irrelevant proportionality constant:
\begin{equation}
n_5^\mathrm{divergent}\sim\mu_5 m_0^2\left(
\log\frac{\Lambda_\mathrm{UV}^2}{\Gamma^2}
\right)^{1-2d_m},\label{eq:approx}
\end{equation}
where $d_m$ is the anomalous mass dimension~\eq{eq:d:m}. In a two-flavor QCD the power of the logarithm in Eq.~\eq{eq:approx} is a small, but positive number: $1-2d_m = 5/29 \approx 0.17$.

\subsubsection{\corr{Zero-point origin of the divergence}}

\corr{
The divergence~\eqref{eq:approx} of the chiral density~$n_5$ occurs if and only if the current quark mass is nonzero, $m_0 \neq 0$. In this section we demonstrate that this divergence has a ``vacuum'' origin rather than a thermodynamic one. 
}

\corr{
The presence of the chiral chemical potential $\mu_5$ modifies the functional behaviour of the energy of the fermionic modes~\eq{eq:E2:pm}, thus affecting the contribution to the often-neglected part of the free energy which would normally be associated with the vacuum energy. Consider, for example, the simplest case of free fermions with the quark mass $m_0$. The positive-energy branch of the fermionic modes in the presence of the nonzero chiral potential $\mu_5$ has the following form:
\begin{equation}
\varepsilon^{(\chi)}_{\bs p}(\mu_5) = \sqrt{(|\bs p| - \chi \mu_5)^2 + m_0^2},
\label{eq:omega:free}
\end{equation}
where $\chi = \pm 1$ labels the helicity of the mode.
}

\corr{
As the chiral chemical potential modifies the spectrum of free field fluctuations, it should also modify the vacuum energy carried by these fluctuations. According to the standard rules of quantum field theory, this zero-point (ZP) contribution is given by the sum over all modes:
\begin{equation}
\Omega_{\mathrm{ZP}}(\mu_5) = \sum_{\chi = \pm 1} \int \frac{d^3 p}{(2 \pi)^3} \varepsilon^{(\chi)}_{\bs p} (\mu_5).
\label{eq:Omega:ZP}
\end{equation}
In all other circumstances, this zero-point contribution is automatically neglected because it does not depend on physical parameters of the system such as temperature and chemical potential and, therefore, may be undoubtedly associated with the vacuum. 
}

\corr{
However in our case, the zero-point energy~\eq{eq:Omega:ZP} depends on the chiral chemical potential~$\mu_5$ and, consequently, the free-energy term $\Omega_{\mathrm{ZP}}$ must be taken into account in additional to the conventional free energy of the system. The explicit dependence of the energy of the zero-point fluctuations~\eq{eq:Omega:ZP} on the chiral chemical potential determines its nonzero contribution to the chiral charge density. In particular, the zero-point energy determines the chiral density at zero temperature, when the usual thermodynamic contribution vanishes:
\begin{eqnarray}
& & n_5{\biggl |}_{T=0} = - \frac{\partial \Omega_{\mathrm{ZP}}}{\partial \mu_5}  
\label{eq:n5:free:m}\\
& & \qquad =  \sum_{\chi = \pm 1} \int_0^\infty \frac{p^2 d p}{\pi^2} 
\frac{(\mu_5 - \chi |\bs p|)}{\sqrt{(\mu_5 - \chi |\bs p|)^2 + m_0^2}},
\nonumber
\end{eqnarray}
which coincides with the functional form of Eqs.~\eq{eq:nba2} and \eq{eq:aray_n5_7} obtained earlier in another way.
}

\corr{
For strictly massless fermions, $m_0 = 0$, the expression under the integral~\eq{eq:n5:free:m} coincides with the hard Fermi cutoff, $\theta(\mu_5 - |{\bs p}|)$, which usually appears in the thermodynamic (and not in the zero-point) part:
\begin{equation}
n_5{\biggl|}_{{}^{m_0 = 0}_{T=0}} \!= \frac{1}{\pi^2} \int_0^\infty p^2 d p\, \theta(\mu_5 - |{\bs p}|) = \frac{\mu_5^3}{3\pi^2}.\
\end{equation}
We automatically recover the finite thermodynamic expression~\eq{eq:array_n5} for $N_f = N_c = 1$.
}

\corr{
As we already figured out, the behavior of the chiral density~\eq{eq:n5:free:m} changes qualitatively for massive fermions with $m_0\neq 0$. At a large momentum $|{\bs p}| \gg m,|\mu_5|$, the integral in Eq.~\eq{eq:n5:free:m} behaves as
\begin{equation}
n_5(T=0) = \frac{\mu_5 m_0^2}{\pi^2} \int^{\Lambda_{\mathrm{UV}}} \frac{d p}{p} \sim \frac{\mu_5 m_0^2}{\pi^2} \log \frac{\Lambda_{\mathrm{UV}}}{m_0} + \dots,
\label{eq:n5:div}
\end{equation}
which leads precisely to the logarithmic divergence found already in Eq.~\eq{eq:approx} (with vanishing anomalous dimension, $d_m = 0$, as it is appropriate to a free theory).
}

\corr{
Evidently, this zero-point divergence of the chiral density~\eq{eq:n5:div} cannot be renormalized by a conventional subtraction method that is usually used with
respect to the vacuum contribution. Thus, we need to modify the normalization prescription itself, in order to deal with a finite chiral chemical potential in 
the theories with a finite fermion mass. We will discuss this question in the next section.
}

\corr{
Before finishing this section, we would like to make three comments.
}

First, this type of divergence does not occur in dense fermionic systems with vanishing chiral chemical potential. Indeed, the vector chemical potential $\mu$ shifts the particle energy modes linearly without modifying the functional form of the momentum dependence of the energy: $\sqrt{|{\bs p}^2| + m_0^2} \to \sqrt{|{\bs p}^2| + m_0^2} \pm \mu$. Therefore, the presence of the vector chemical potential does not affect the zero-point energy. 

Second, in a local NJL model, in which the constituent 
quark mass $M$ has no momentum dependence, we would get the constituent quark
mass $M$ instead of the current quark mass $m_0$ in Eq.~\eqref{eq:approx}
which would give a divergence in the chiral limit as well. This is precisely the
divergence that has been found in Ref.~\cite{Gatto:2011wc}.
On the other hand, within the nonlocal NJL model, if we were in the chiral limit
we would have gotten a finite  $n_5$.  \corr{The divergence, as we have just demonstrated, 
appears due to the zero-point fluctuations which exist in both local and non-local models.}

Third, from the context of this discussion, it is difficult to say \corr{with definiteness} if this divergence
of the chiral density $n_5$, in particular Eq.~\eqref{eq:approx}, is valid also in full QCD:
it is true that the present model represents a very crude description of the interaction that leads
to the spontaneous chiral symmetry breaking of QCD. However, 
\corr{the model} describes correctly the behavior of the quark mass function
at large $p_E$ which is the region of interest for the discussion of the divergence of $n_5$,
giving 
\corr{an argument}
that the result in Eq.~\eqref{eq:approx} 
\corr{is applicable}
to QCD as well.

\subsubsection{Renormalization of the chiral chemical potential}

The chiral density has been computed in the first-principle numerical calculations in the scope of lattice QCD endowed with the dynamical rooted staggered fermions~\cite{Astrakhantsev:2019wnp}. To this end, the Lagrangian has been shifted by a lattice version of the source term $\mu_5 n_5$ 
which effectively induces a nonzero the chiral density $n_5 \neq 0$ at a nonvanishing chiral chemical potential $\mu_5 \neq 0$.
While the UV divergence of the chiral density $n_5$ in the linear term has been also been noticed for naive formulation of lattice fermions, the full dynamical QCD calculations seem to support a finite result $n_5 \sim \Lambda_{\mathrm{QCD}}^2 \mu_5$, where $\Lambda_{\mathrm{QCD}}$ is a finite mass parameter of the order of a typical QCD energy scale.  This result matches well the Chiral Perturbation Theory ($\chi$PT) which implies $n_5\sim f_\pi^2 \mu_5$ for small $\mu_5$ at $T=0$ in the chiral limit~\cite{Astrakhantsev:2019wnp}. Here $f_\pi \sim  \Lambda_{\mathrm{QCD}}$ is a pion decay constant. It is an easy exercise to prove that Eq.~\eqref{eq:n5_2} is consistent with $\chi$PT at the lowest order in the chiral chemical potential $\mu_5$ and in the current quark mass by putting $\mu_5=0$ in the integrand and using Eq.~(4.26) of Ref.~\cite{Klevansky:1992qe}.

We want to notice, however, that we have been unable to reproduce the $\chi$PT relation $n_5\sim f_\pi^2 \mu_5$ for a non-regularized chiral density $n_5$
in the context of the nonlocal NJL model. This relation is unlikely to hold in this model because the integrals for the naively-defined $n_5$ and the pion decay constant, $f_\pi$, entering at different sides of the relationship, are pretty inconsistent (see, for example, Ref.~\cite{Hell:2008cc}). In particular, $f_\pi$ is given in terms of convergent integrals, while the non-regularized $n_5$ still possesses a UV divergence in this model~\eq{eq:approx}.
In the nonlocal NJL model, we can renormalize the chiral condensate~$n_5$ by removing the logarithmic divergence~\eq{eq:approx}
similarly to a standard renormalization procedure in any well-defined renormalizable theory. 

\corr{
As we discussed in the previous section, the divergence appears from zero-point fluctuations of the fermionic modes which possess the unconventional energy dispersion in the presence of the chiral chemical potential. This feature appears both in the non-local NJL model~\eq{eq:E2:pm} as well as in the case of free fermions~\eq{eq:omega:free}. In the standard quantum field theory, the contribution from the zero-point fluctuations can easily be subtracted because it does not depend on the parameters of the matter sector of the theory. In our case, however, the zero-point term contains also a matter contribution which cannot be neglected. This leads to subtleties in the normalization procedure.
}

Postponing the physical justification of the formal renormalization procedure to a later discussion, we notice that 
the source of the \corr{zero-point ultraviolet} divergence can be easily traced in Eq.~\eqref{eq:integranda_3}. 
We define a renormalized chiral density as follows:
\begin{equation}
n_5^\mathrm{R} =4 N_c N_f\mu_5 \left( {\cal H}-{\cal H}_0\right),
\label{eq:ren_n5_1}
\end{equation}
where the counterterm 
\begin{equation}
{\cal H}_0 =  \int\frac{d^4p_E}{(2\pi)^4}
\frac{(3\bm p^2 - p_4^2) m(p_E)^2}
{[p_E^2 + m(p_E)^2]^3}.
\label{eq:H0_def}
\end{equation}
is computed at zero temperature $T=0$ and it is independent on the chiral condensate. 

When multiplied by 
the chiral chemical potential $\mu_5$ in Eq.~\eqref{eq:ren_n5_1}, the regularized contribution depends only on $\mu_5$. 
The term $m(p_E)^2$ that appears in the denominator of the integrand in Eq.~\eqref{eq:H0_def} makes it possible to avoid 
the apparent infrared divergence at low Euclidean momenta, $p_E \to 0$. This definition of the chiral density is consistent 
with Eq.~\eqref{eq:array_n5} in the chiral limit, at which both the constituent and current masses are zero, $M=m=0$.

The subtraction in Eq.~\eqref{eq:ren_n5_1} is enough to cancel the mild logarithmic divergence of the chiral density~\eqref{eq:approx}
since higher-order terms in powers of the chemical potential $\mu_5$ in Eq.~\eqref{eq:integranda_3} would lead to convergent integrals. Moreover, 
we also verified that within the nonlocal NJL model at zero temperature $n_5\simeq C_M M_q^2 \mu_5$, where $M_q$ denotes
the quark mass function at $p=0$ and finite $\mu_5$. Numerically, the proportionality constant turns out to be $C_M \simeq 0.5$ at low chemical potentials
up to $\mu_5 \approx 300$ MeV, while for larger chemical potentials $\mu_5$ the proportionality becomes softer, finally reaching $C_M \simeq 0.3$ 
at $\mu_5=400$ MeV. These results are in agreement with the lattice QCD studies reported in Ref.~\cite{Astrakhantsev:2019wnp} 
in which it is found that the chiral density $n_5$ at zero temperature is of the order of the anticipated value $\mu_5\Lambda_\mathrm{QCD}^2$.

The straightforwardly-defined chiral density is finite in the chiral limit (where the fermion mass is zero) and divergent for the massive fermions~\eqref{eq:approx}. Therefore, it is natural to suggest that the divergence of the chiral density~\eq{eq:approx} is directed related to the presence of the fermions' mass, and, naturally, to the inconsistency of the notion of the chiral chemical potential for fermions with a nonzero mass. Indeed, the chiral charge is not conserved as it dissolves via the chirality flips for a massive fermion regardless of the origin of it mass which could be either a current mass or a dynamically-generated mass. For a free Dirac fermion with a mass $M$, the axial (chiral) current~\eq{eq:j:5} has a nonzero 4-divergence at the level of the classical equations of motion:
\begin{equation}
\partial_\mu j^\mu_5 = 2 i M {\bar \psi} \gamma^5 \psi.
\label{eq:nonconservation}
\end{equation}
For strictly massless fermions, the chirality is a conserved number: $\partial_\mu j^\mu_5 = 0$.

A generic chemical potential has a well-defined meaning only for conserved quantities such as the electric (vector) charge. One may alternatively say, that no chemical potential can thermodynamically be conjugated to a non-conserved quantity. In the photodynamics, for example, it is impossible to self-consistently introduce a chemical potential for the total number of photons as the latter number is evidently not conserved. In a free theory, such a ``chemical potential'', associated with a non-conserved quantity, decouples from the dynamics of the theory. In an interacting theory, this type of chemical potential may affect the dynamics as we discuss below.

In QCD, the chiral properties of fermions are connected to the topology of the gluonic sector of the theory. For example, the instanton- and sphaleron-induced transitions between adjacent topologically distinct vacua induce changes in the chiral charge of the fermions due to the axial anomaly in QCD:
\begin{equation}
\left( N_R - N_L \right){\Bigl |}_{t \to - \infty} - \left( N_R - N_L \right) {\Bigl |}_{t \to + \infty}  = - 2 N_f \Delta N_{\mathrm{CS}},
\end{equation}
where $N_5 \equiv V n_5 = N_R - N_L$ is the difference between the numbers of fermions possessing the right-handed ($N_R$) and left-handed ($N_L$) chiralities, and change $\Delta N_{\mathrm{CS}}$ in the Chern-Simons number between initial and final configurations is given by the topological charge~\eq{eq:CS} of the gluonic configuration.

Since the fermion and gluon sectors of the theory are interacting with each other, a finite chiral chemical potential may also induce a response in the topological gluon sector in thermal equilibrium of QCD. Therefore, the finite chiral chemical potential may have a physical sense in QCD with massive fermions, regardless of the fact that the chiral number is not a conserved quantity. The relaxation of the chiral density may induce physical changes in the topological charge fluctuations in the gluonic sector of the theory. We will see below that the topological susceptibility of the theory is seemingly correlated with the chiral density in the background of the chiral chemical potential.  

Since the chiral charge is not conserved~\eq{eq:nonconservation}, the introduction of the chiral chemical potential~\eq{eq:Lq} cannot be justified in the thermodynamical sense. On the contrary, the chiral chemical potential $\mu_5$ should be treated as a new coupling of QCD, which -- in general -- needs a renormalization. The corresponding thermodynamically conjugated quantity, the density, should also require a renormalization as well (for example, in QCD, both the gluon coupling and gluon fields/strength tensors are renormalized perturbatively). Therefore, Eq.~\eq{eq:ren_n5_1} defines nothing but a certain renormalization scheme in the chiral sector.

\section{Results}

In this section we report the main results of our study on the nonperturbative quantities associated with the chiral quark density in the scope of the nonlocal NJL model. Below we discuss the dynamical quark mass $M$, the chiral condensate $\langle  {\bar q} q\rangle$, the chiral density $n_5$, the topological susceptibility $\chi_{\mathrm{top}}$, and, finally, the phase diagram in the $(\mu_5,T)$ parameter plane as originates from the nearly-critical behaviour of the chiral density and the topological susceptibility.

First, we fix the parameters of the NJL model. We take $\Lambda=550$ MeV for the matching scale, $m_0=5$ MeV for the undressed quark mass. 
Finally, the value of quartic quark interaction constant $G=2.6/\Lambda^2$ and $G_1=(1-c)G_1$ is chosen in order to
reproduce the phenomenological value of the light quark condensate $\langle\bar qq\rangle = (-250~\mathrm{MeV})^3$ in the vacuum at $T=\mu_5 = 0$, and $G_2=cG$ with $c=0.2$ \cite{Lu:2018ukl}.
In the numerical computations we  regularize the thermodynamic potential~\eqref{eq:omega_1} by summing over all the Matsubara's frequencies
and restricting the integration over the 3-momentum at the momentum sphere $|{\bs p}| \leq\Lambda_\mathrm{UV}$ with the ultraviolet cutoff $\Lambda_\mathrm{UV} = 3\Lambda$ (we have verified that changing this UV cut does not change the results drastically).

\subsection{Catalysis of chiral symmetry breaking}

Our first aim is demonstrate that the nonlocal NJL model is capable to describe the catalysis of chiral symmetry breaking induced by $\mu_5$, at least for relatively moderate values of the latter, $|\mu_5| \lesssim \Lambda$. To this end, we first discuss the constituent quark mass and the quark condensate at finite temperature $T$ and chiral chemical potential $\mu_5$. As some features of the quark mass at $\mu_5 \neq 0$ were already been discussed in Refs.~\cite{Ruggieri:2016ejz,Frasca:2016rsi}, here we limit ourselves to show the evolution of the quark mass at zero Euclidean momentum $p_E=0$ with varying $T$ and $\mu_5$. For the computation of the quark mass, we restrict ourselves to vanishing topological angle $\theta=0$. Therefore, the pseudoscalar condensate vanishes, $\eta=0$, while the chiral condensate $\sigma$ is the only condensate left.

\subsubsection{Constituent quark mass}

\begin{figure}[t!]
\begin{center}
\includegraphics[scale=0.6]{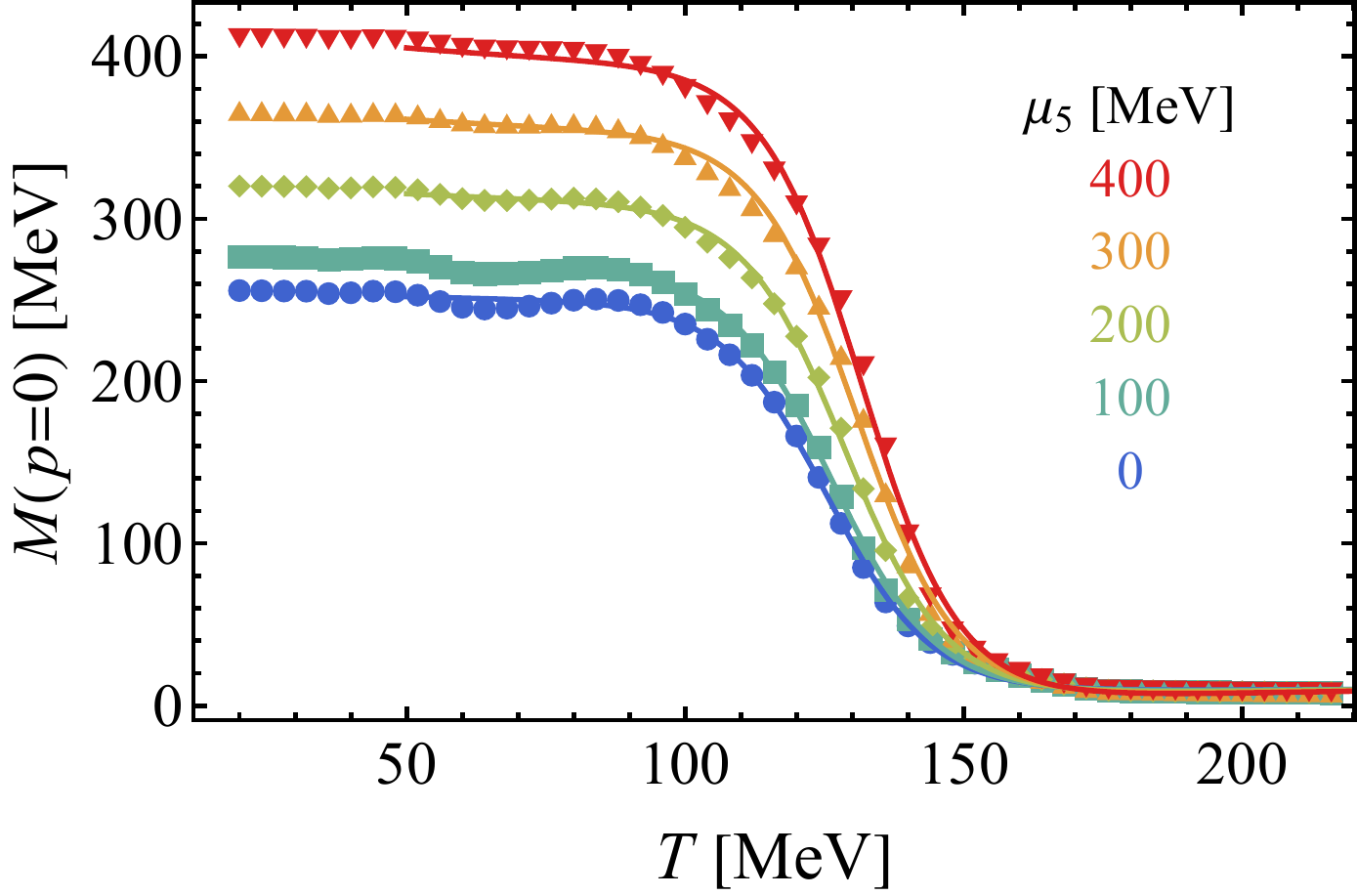}
\end{center}
\caption{\label{Fig:masses}Constituent quark masses $M$ at a zero momentum $p=0$ versus temperature $T$.
The blue dots denote a vanishing chiral chemical potential $\mu_5=0$,
the green squares stand for $\mu_5=100$ MeV, the light green diamonds correspond to $\mu_5=200$ MeV, the
orange upwards-pointing  triangles stand for $\mu_5=300$ MeV, and, finally,  
the red downwards-pointing triangles denote $\mu_5=400$ MeV. The solid lines represent the best fits~\eq{eq:tanh:fit}.}
\end{figure}

In Fig.~\ref{Fig:masses} we plot the constituent quark mass at zero Euclidean momentum versus temperature. 
The figure indicates that the increase in the chiral misbalance enhances the chiral symmetry breaking at any temperature
in agreement with conclusions of Ref.~\cite{Braguta:2016aov}.

In our model, we have verified that the catalysis of the chiral symmetry breaking occurs up to $\mu_5\approx\Lambda$. 
Above this value, the constituent mass decreases with $\mu_5$, but this feature might be
related to the fact that we have not included any backreaction on the interaction kernel.
We also notice that the constituent mass decreases in a narrow range of temperatures: there is a crossover to a high temperature
phase in which the chiral symmetry is approximately restored. The smoothness of the transition appears to be supported by thermal fluctuations.
We will see in the next section that the chiral symmetry restoration is accompanied by the approximate restoration of the $U(1)_A$ symmetry as well.

One way to determine position of the thermal crossover is to identify it, at a fixed chemical potential $\mu_5$, with the temperature $T_c = T_c(\mu_5)$ at which the absolute value of the slope of the infrared quark mass $|dM(p_E=0)/dT|$ takes its maximum. For example, one finds $T_c \simeq 125$ MeV at $\mu_5=0$ which agrees very well with the value quoted earlier in Ref.~\cite{Pagura:2016pwr}. Figure~\ref{Fig:masses} also implies that the critical temperature is an increasing function of $\mu_5$ in the whole range of chemical potentials $\mu_5$ covered by this study. 

A more accurate way to find the thermal crossover is to find a suitable fitting function that may smoothly interpolate the low- and high-temperature behaviour of the constituent quark mass. And indeed, this quantity may be well described in a wide region of temperatures by the following function (with ${{\mathcal O}} = M$ in the considered case):
\begin{equation}
{\mathcal O}(\mu_5,T) = C_1(\mu_5) T^\nu \tanh \left( \frac{T - T_c(\mu_5)}{\delta T_c(\mu_5)}\right) + C_2(\mu_5),
\label{eq:tanh:fit}
\end{equation}
where $C_1$, $C_2$, $\nu$, $T_c$, and $\delta T_c$ are the fitting parameters defined at each fixed value of the chiral chemical potential~$\mu_5$. The power $\nu$ is usually quite small ($\nu \sim 0.1$ or smaller). Below, we will use the generic function~\eq{eq:tanh:fit} to describe other quantities in the pseudocritical region. We do not put the superscript $\mathcal O$ to the fitting parameters to keep our notations concise. 

The best fits of the mass gap $M$ by the function~\eq{eq:tanh:fit} are shown in Fig.~\ref{Fig:masses} by the solid lines.  The pseudocritical temperature $T_c = T_c(\mu_5)$ and the width $\delta T_c = \delta T_c (\mu_5)$ the pseudocritical region, as determined by the mass gap $M \equiv M(p=0)$, follow very closely the corresponding quantities for the chiral condensate and the topological susceptibility, that we will discuss in more detail below.

\subsubsection{Chiral condensate}

\begin{figure}[t!]
\begin{center}
\includegraphics[scale=0.6]{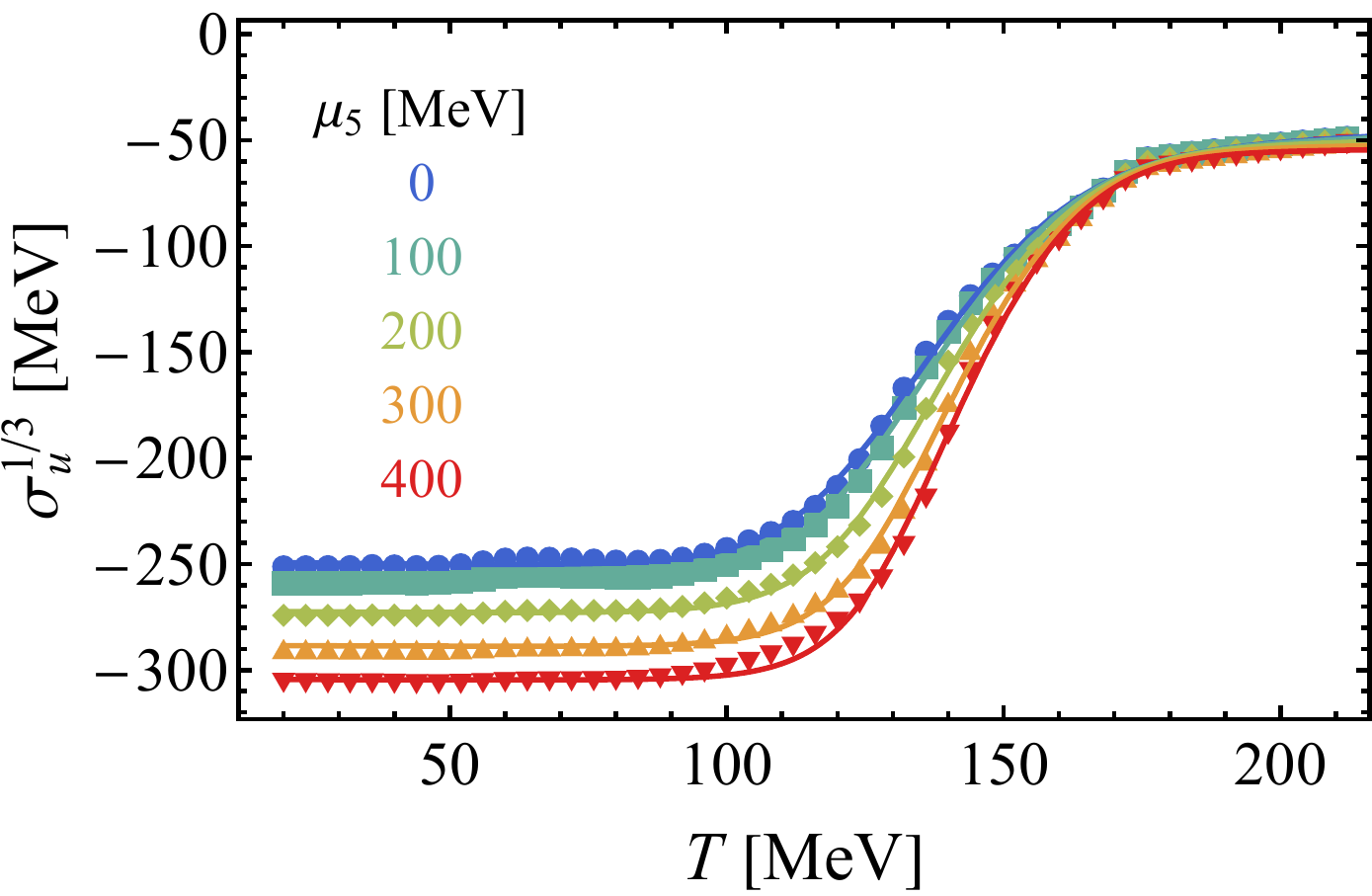}
\end{center}
\caption{\label{Fig:ccv}Chiral condensate $\sigma_u\equiv\langle\bar u u\rangle$ versus temperature.
Conventions for colors and symbols are the same as in Fig.~\ref{Fig:masses}. The solid lines represent the best fits~\eq{eq:tanh:fit}.}
\end{figure}

The catalysis of the chiral symmetry breaking, induced by the chiral chemical potential $\mu_5$, is also evident from the behaviour 
of the chiral condensate, $\langle\bar qq\rangle$, as
defined in Eq.~\eqref{eq:cc1}. In Fig.~\ref{Fig:ccv} we plot the chiral condensate for one of the light quarks, $\sigma_u\equiv\langle\bar u u\rangle$,
versus temperature for several fixed values of $\mu_5$. 
The magnitude of the chiral condensate increases with the rise in the chiral chemical potential 
$\mu_5$. Thus, we have yet another confirmation that the chiral chemical potential 
acts as a catalyzer of the chiral symmetry breaking.

The approximate restoration of the chiral symmetry at a finite temperature via a smooth crossover and not via a 
real thermodynamic phase transition. In the absence of a thermodynamic singularity in the parameter space, the 
very definition of the critical temperature leaves a large room for ambiguities. Moreover, the transition temperature 
depends not only on the method of its definition, but also on the particular thermodynamic quantity used to identify 
the temperature. The chiral condensate  $\sigma_u\equiv\langle\bar u u\rangle$ (and, equivalently, for $\sigma_d 
\equiv \sigma_u$) may be described by the same type of function that has also been used for the constituent quark~\eq{eq:tanh:fit}.
The best fits are shown in Fig.~\ref{Fig:masses} by the solid lines. We will discuss the pseudocritical temperature $T_c = T_c(\mu_5)$ 
and the width $\delta T_c = \delta T_c (\mu_5)$ in more detail at the end of this section.

\subsection{Chiral density}

\begin{figure}[t!]
\begin{center}
\includegraphics[scale=0.56]{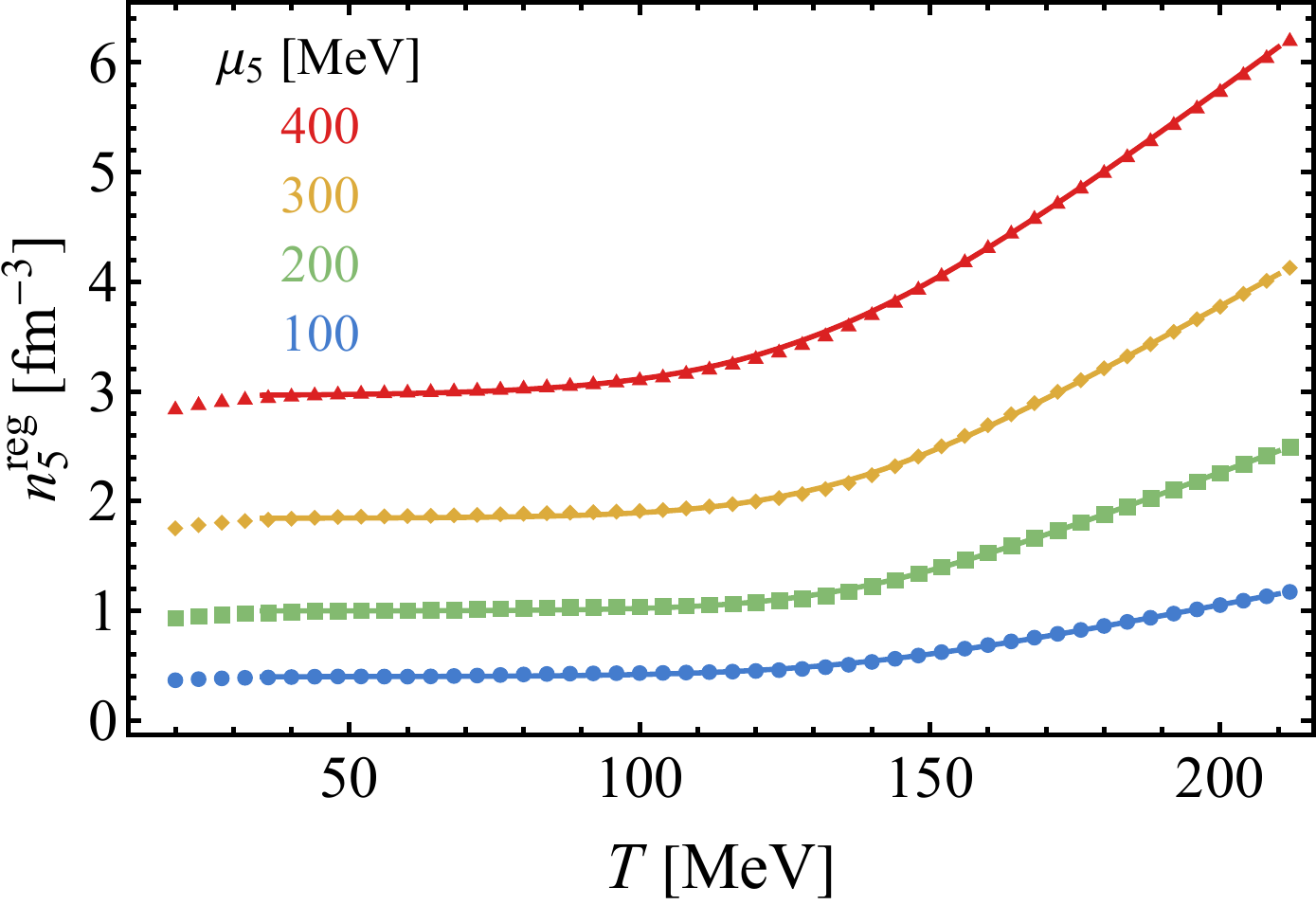}
\end{center}
\caption{\label{Fig:n5}Regularized chiral density~$n_5$, defined via Eqs.~\eq{eq:ren_n5_1} and \eq{eq:H0_def}, versus temperature.
Conventions for colors and symbols are the same used in Fig.~\ref{Fig:masses}. The solid lines show the fits by the function~\eq{eq:dens:fit}.}
\end{figure}

In Fig.~\ref{Fig:n5} we plot the regularized chiral density~$n_5$, defined in Eqs.~\eq{eq:ren_n5_1} and \eq{eq:H0_def}, versus temperature $T$ for several values 
of the chiral chemical potential~$\mu_5$. The qualitative trend of $n_5$ versus temperature is obviously the same for every fixed $\mu_5$: as temperature rises, 
the chiral density increases. This property is related to the fact that thermal excitations contribute more to the thermodynamic potential at temperature rises.
Alternatively, the increase of the chemical potential $\mu_5$ at any fixed temperature $T$ results in the increase of the chiral density.

For every fixed chiral chemical potential $\mu_5$, the chiral density exhibits a knee-like structure which separates the low-temperature from high-temperature behaviour, as seen in Fig.~\ref{Fig:n5}. This behavior may be described with a very good accuracy by the following function:
\begin{equation}
n_5(\mu_5,T) = n_1(\mu_5) F \left( \frac{T - T_0(\mu_5)}{\delta T_0(\mu_5)}  \right) + n_2(\mu_5),
\label{eq:dens:fit}
\end{equation}
where $F(x) = \ln \left( 1 + e^x \right)$. The fitting parameters are $n_1$, $n_2$, $T_0$, and $\delta T_0$, where the temperature $T_0$ discriminates between the low- and high-temperature behaviour while the quantity $\delta T_0$ has a sense of the width of the transition region. All fitting parameters are the functions of the chiral chemical potential $\mu_5$. The best fits are shown in Fig.~\ref{Fig:n5} by the solid lines.

Since the chiral density $n_5$ is not an order parameter of the deconfining transition, the knee temperature $T_0$ does not have a meaning of a (pseudo) critical temperature, especially, for a crossover transition. We get $T_0 \simeq 135\,{\mathrm{MeV}}$ with the width $\delta T_0 \simeq 20\,{\mathrm{MeV}}$.

\subsection{Topological susceptibility}

The topological susceptibility measures the strength of fluctuations of the topological charge in the medium.
In our model, the topological susceptibility can be computed as the curvature:
\begin{equation}
\chi_\mathrm{top} = \left.\frac{\partial^2\Omega}{\partial\theta^2}\right|_{\theta=0}.
\label{eq:chitop1}
\end{equation}

It is well known that for two degenerate flavors of light quarks, the topological susceptibility~\eq{eq:chitop1} is related to the quark condensate $|\langle\bar q q\rangle|$ 
follows: $\chi_\mathrm{top} = m|\langle\bar q q\rangle|$ where $m$ is the current quark mass. This relation, valid at zero temperature, shows the link between
the fluctuations of the topological charge~\eq{eq:chitop1} and the dynamics of the light quark flavors that leads to the spontaneous chiral symmetry breaking. 
First-principle lattice simulations~\cite{Gattringer:2002mr,Cossu:2013uua,Ding:2017giu} indicate that the topological susceptibility tends to decrease with rising temperatures and that the axial symmetry tends to be restored at high temperatures at the chirally unbroken phase.

In a local NJL model, the topological susceptibility with light quarks in the chiral medium has been studied for the first time in Ref.~\cite{Gatto:2011wc}. 
It has also been recently subjected to the first-principle lattice calculations at zero temperature in Ref.~\cite{Astrakhantsev:2019wnp}.
Both studies agree qualitatively with each other on the fact that at zero temperature $\chi_\mathrm{top}$ increases with $\mu_5$,
while a quantitative comparison is not feasible due to the different quark masses used in these calculations.

On the other hand, the local NJL model~\cite{Gatto:2011wc} indicates that when the temperature is close to $T_c$, 
the chiral chemical potential tends to lower $\chi_\mathrm{top}$. This property is an artifact of the local interaction and of the 3-dimensional regulator
used in Ref.~\cite{Gatto:2011wc}. Already with a 4-dimensional regulator, the critical temperature $T_c$ increases with increase 
of the chiral chemical potential $\mu_5$, at least for small $\mu_5$ \cite{Ruggieri:2016ejz}. Cutoff effects become substantial 
at large values of $\mu_5$ in any regularization scheme and $T_c$ tends to be lowered with the rise of $\mu_5$. For this reason, here we 
compute $\chi_\mathrm{top}$ both at zero and at finite temperature by using the nonlocal NJL model, which offers a more trustable 
response of critical temperature $T_c$ considered as the function of the chiral chemical potential $\mu_5$.

\begin{figure}[t!]
\begin{center}
\includegraphics[scale=0.6]{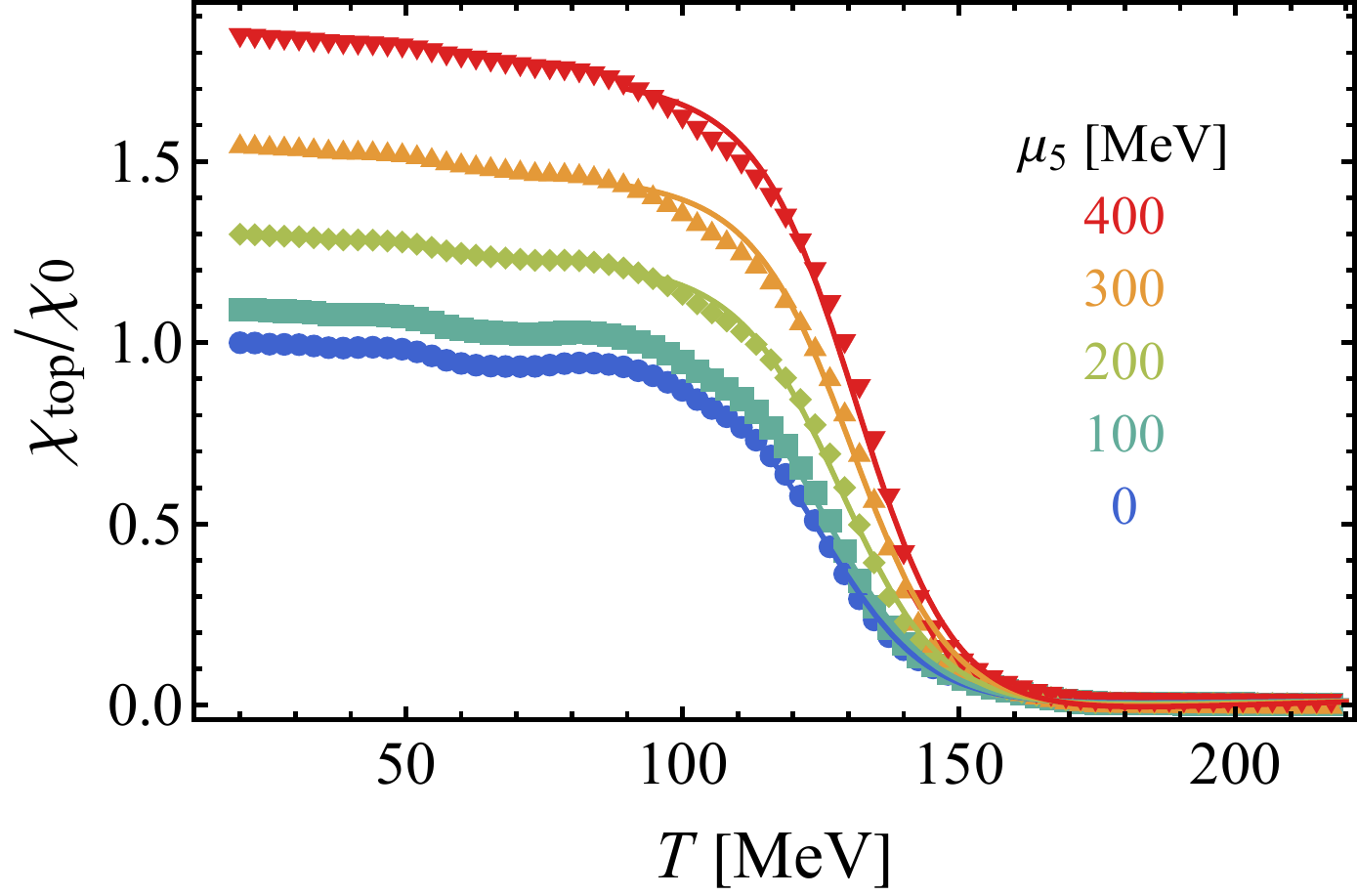}
\end{center}
\caption{\label{Fig:ctop}Topological susceptibility~\eq{eq:chitop1} versus temperature, in units of the vacuum susceptibility~$\chi_0$~\eq{eq:chi:0}.
Conventions for colors and symbols are the same used in Fig.~\ref{Fig:masses}. The solid lines represent the best fits by the function~\eq{eq:tanh:fit}.}
\end{figure}

In Fig.~\ref{Fig:ctop} we plot $\chi_\mathrm{top}$ versus temperature. In the figure the susceptibility is given in units of its vacuum value,
\begin{equation}
\chi_0\equiv \chi_\mathrm{top}{\Bigl|}_{T=\mu_5=0} \simeq 76\,{\mathrm{MeV}},
\label{eq:chi:0}
\end{equation}
where the numerical number is presented for the set of parameters used in our study.

We notice that the increase the chiral density results in the increase of the topological susceptivity at each fixed temperature. 
This result is in a disagreement with previous calculations that use the local NJL model~\cite{Gatto:2011wc},
which instead predict an increase of $\chi_\mathrm{top}$ at small $T$ but a decrease of $\chi_\mathrm{top}$ at large $T$.
The result of \cite{Gatto:2011wc} was obtained within a local NJL model, thus the behavior of $\chi_\mathrm{top}$ in that model
follows that of the constituent quark mass at finite $\mu_5$.

Figure~\ref{Fig:ctop} indicates that there is a narrow range of temperatures in which the susceptibility $\chi_\mathrm{top}$ decreases abruptly thus signaling the partial restoration of the axial $U(1)_A$ symmetry. In order to extract the pseudocritical temperature more accurately, we describe the topological susceptibility by the function~\eq{eq:tanh:fit}.
The best fits are shown in Fig.~\ref{Fig:ctop} by the solid lines.  In the next subsection, 
we will discuss the pseudocritical temperature $T_c$ for this topological crossover in more detail.

\subsection{Pseudocritical temperatures}

\begin{figure}[t!]
\begin{center}
\includegraphics[scale=0.6]{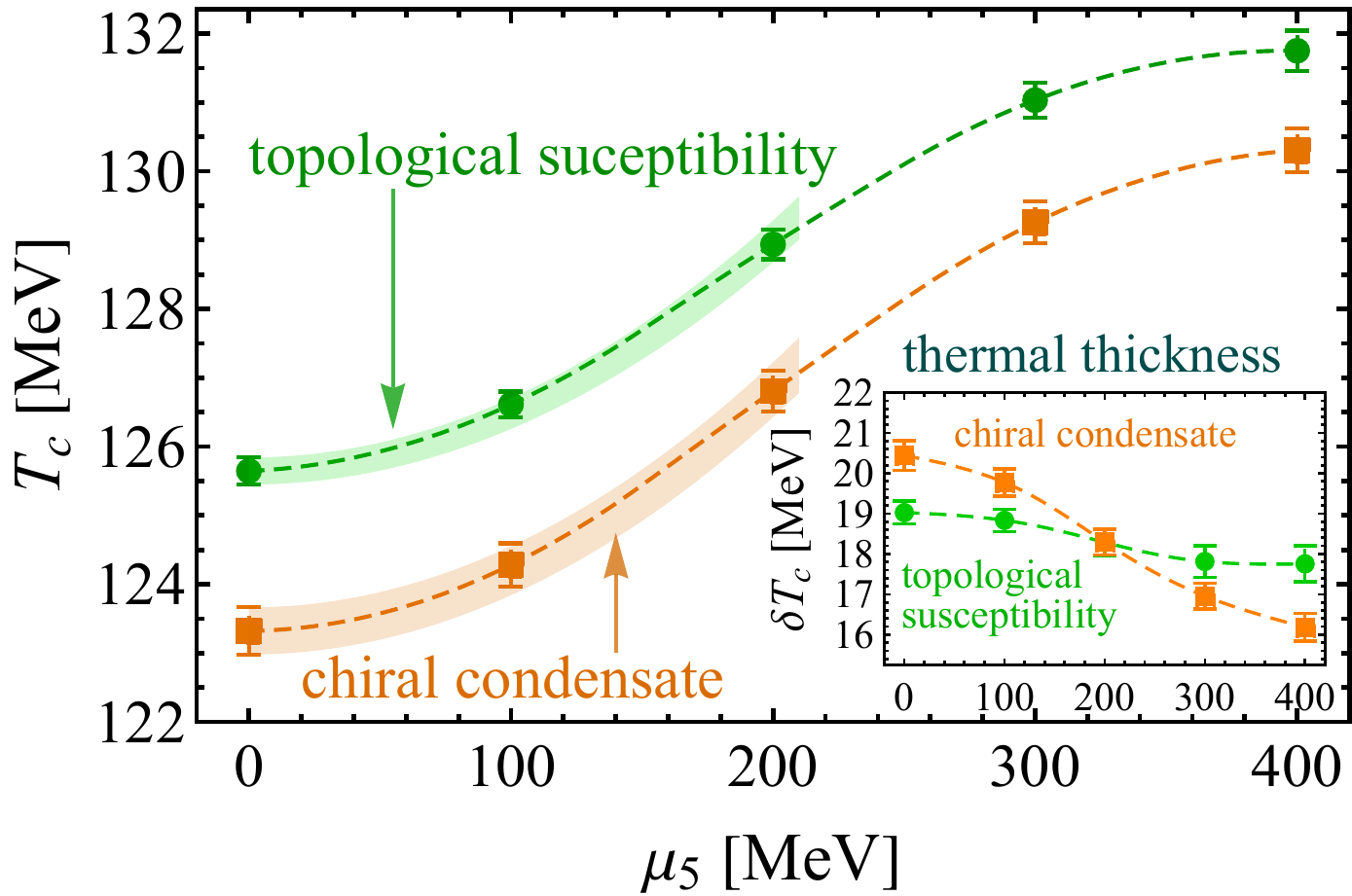}
\end{center}
\caption{\label{Fig:Tc_mu5} Pseudo-critical temperatures $T_c$ of the chiral symmetry restoration (the orange squares) and the axial symmetry restoration (the green circles) as functions of the chiral chemical potentials $\mu_5$. The inset shows the thickness of each transition. The data are obtained by fits of, respectively, the chiral condensate (Figs.~\ref{Fig:ccv}) and the topological susceptibility (Fig.~\ref{Fig:ctop}), see the text tor more details. The dashed lines are drawn to guide the eye. The shaded regions represent the quadratic-curvature behavior of Eq.~\eq{eq:kappa:def}.}
\end{figure}

In Figure~\ref{Fig:Tc_mu5} we compare the pseudo-critical temperatures $T_c$ of the chiral and axial crossovers, as well as their thermal widths, as functions of the chiral chemical potential~$\mu_5$. 

The figure shows a few interesting qualitative features of the chiral phase transition. First of all, both pseudocritical temperatures rise in the unison as the chiral density increases. Second, the restoration of the axial symmetry, as revealed by the topological susceptibility, appears at a higher temperature than the restoration of the chiral symmetry. This statement is independent of the chiral chemical potential. At large values of the chiral chemical potential we observe a flattening of the both pseudocritical temperatures which, however, should be attributed to the crudeness of the model used in our studies. In particular, we have neglected any possible backreaction induced 
by the chemical potential $\mu_5$ on the form factor. From this point of view, it would be interesting the computation of the quark mass function at finite $\mu_5$ using 
the Schwinger-Dyson equation, which should allow to incorporate the aforementioned dependence in the mass formfactor.

The widths of both axial and chiral crossover transitions, $\delta T_c$, show a tendency for shrinking as the chiral matter gets denser, thus implying a strengthening of both transitions as the chiral chemical potential increases. However, we see no signature of a critical endpoint in the $(\mu_5,T)$ phase diagram in the studied region of the chiral chemical potential $\mu_5$.

These conclusions are in a certain disagreement with previous studies that used a local interaction 
kernel~\cite{Chernodub:2011fr,Ruggieri:2011xc}
as well as with recent works using the Wigner function technique~\cite{Das:2019crc}. 
On the other hand, the absence of the phase transition point in the $(\mu_5,T)$
plane agrees well with the results obtained within the nonlocal NJL models~\cite{Ruggieri:2016cbq,Ruggieri:2016ejz,Ruggieri:2016xww,Frasca:2016rsi}, with the solution of the Schwinger-Dyson equations \cite{Xu:2015vna,Wang:2015tia}, 
and with the first-principle QCD studies in a limited range of temperatures and chemical potentials~\cite{Braguta:2015owi}.
  
A quantitative analysis reveals a less sharp picture given the smooth nature of both crossover transitions. First of all, the magnitude of the increase of both axial and chiral pseudo-critical temperatures is very small: the rise of the chemical potential, from $\mu_5 = 0$ to rather large value $\mu_5 = 400\,\mathrm{MeV}$, leads to the enhancement of the critical temperature by about $\Delta T_c \simeq 5\,\mathrm{MeV}$, or less than by $5\%$. This variation of the temperature is located well within the broad widths of both crossovers, which are wider than $15\,\mathrm{MeV}$ for all studied values of the chiral chemical potential. Moreover, at each given chiral density, both crossovers overlap strongly since the difference in the axial and chiral pseudocritical temperatures is within $(2{-}3)\,\mathrm{MeV}$.

We also determine the curvature $\kappa_5$ of the crossover transition in the $(\mu_5,T)$ plane:
\begin{equation}
\frac{T_c(\mu_5)}{T_c(0)} = 1 - \kappa_5 \frac{\mu^2_5}{T_c^2(0)} + \dots,
\label{eq:kappa:def}
\end{equation}
which is usually applied to the low-density domain with $|\mu_5| \ll T_c$. According to the conventions used in the literature, a positive curvature $\kappa$ corresponds to a diminishing (pseudo-) critical temperature as the chemical potential increases. In a realistic QCD with three quark flavors ($N_f = 2+1$, with two light $u$ and $d$ quarks, and one heavier $s$ quark), the baryonic curvature $\kappa_B$ determines the curvature of the pseudocritical temperature with respect to increase of the baryonic potential $\mu_B = 3 \mu$~(see, for example, the recent studies in Ref.~\cite{Bazavov:2018mes}):
\begin{equation}
\frac{T_c(\mu_5)}{T_c(0)} = 1 - \kappa_B \frac{\mu^2_B}{T_c^2(0)} +\dots ,
\label{eq:kappa:B:def}
\end{equation}
where $\mu \equiv \mu_q$ is the quark chemical potential. In our article, we identify the chiral curvature~\eq{eq:kappa:def} with respect to the chiral (axial) chemical potential $\mu_5 \equiv \mu_A$.

The curvatures of the chiral and axial crossovers for the chiral quark chemical potential approximately coincide and give $\kappa_5^{\mathrm{axial}} = - 0.0105(4)$
obtained from the topological susceptibility and $\kappa_5^{\mathrm{chiral}} = - 0.0108(3)$ as extracted from the chiral condensate. The corresponding quadratic dependences are shown in Fig.~\ref{Fig:Tc_mu5} by the shaded regions. The width of each region corresponds to the statistical error. We would like to notice that the quadratic dependence of the critical temperature on the chiral chemical potential holds very well well for the relatively large values of the chiral chemical potential, $\mu_5 \sim T_c$.

\section{Summary and Conclusions}

We have reported on our study of the chiral and axial symmetry breaking in chirally-imbalanced QCD with two flavors of light fermions at finite temperature using a nonlocal Nambu--Jona-Lasinio model. We studied the chiral condensate, $\langle {\bar q} q \rangle$, the topological susceptibility, $\chi_\mathrm{top}$, and chiral density, $n_5$, of a chiral medium, namely a system with chiral chemical potential $\mu_5\neq0$, at finite temperature.
All the calculations have been performed within a nonlocal NJL model with quark mass function that agrees with perturbative QCD at large
Euclidean momentum. Our approach differs from almost all of the previous calculations in which local effective models have been used.
The use of a nonlocal NJL model is favored over the local ones since the former predicts that the critical temperature for
the approximate chiral symmetry restoration, $T_c$, increases with $\mu_5$ in agreement with the solution of the Schwinger-Dyson equations as well as with the first-principle lattice QCD calculations. On the contrary, the conclusions of our approach disagree with the predictions of local the NJL model as well as with the results obtained recently within the Wigner function approach.

The response of the chiral condensate to a finite chiral charge density at zero and finite temperature
shows that the chiral chemical potential $\mu_5$ serves as a catalyzer of chiral symmetry breaking: 
the chiral condensate strengthens as the chiral density increases.
This conclusion is in agreement with other studies~\cite{Braguta:2016aov}.
Moreover, the behaviour of the topological susceptibility at a finite chiral chemical potential 
indicates that the chiral medium tends to break the axial symmetry as well:
the topological susceptibility becomes larger with increase of the chiral density at all studied temperatures. 
In other words, the chiral medium increases the fluctuations of the topological charge, thus enhancing the breaking of 
the axial $U(1)_A$ symmetry.
We have confirmed that the critical temperature of the chiral crossover rises with $\mu_5$; 
we also noted the same behaviour for the axial crossover and pointed out an apparent hierarchy of the temperatures 
$T_c^{\mathrm{axial}} > T_c^{\mathrm{chiral}}$. However,
the axial and chiral crossovers possess substantial thermal widths, $\delta T_c \sim (15-20)\, \mbox{MeV}$, 
which imply that these transitions overlap as $T_c^{\mathrm{axial}} - T_c^{\mathrm{chiral}} \sim (2-3)\, \mbox{MeV}$. 
Thus, in our model, the axial symmetry restoration happens simultaneously with chiral symmetry restoration.

Part of this study has been devoted to the divergence of the chiral density.
We argued that the presence of the chiral chemical potential should be treated as a Lorentz-frame-dependent coupling. The main argument is that the quarks get substantial masses due the chiral symmetry breaking while the chiral charge is not a classically conserved quantity for the massive fermions. Therefore, the corresponding thermodynamically-conjugated chiral chemical potential should not, therefore, be considered as a true chemical potential. The divergence of an unrenormalized chiral density~\eq{eq:approx} is a consequence of this property. 

\corr{
Technically, the presence of a nonzero chiral chemical potential modifies the functional form of the momentum dependence of the fermionic eigenenergies. The latter contributes to the zero-point energy which is no more associated with the pure vacuum contribution due to the apparent dependence of the chiral chemical potential. Consequently, the zero-point fermionic fluctuations contribute to the density of the chiral charge. The zero-point contribution is finite for massless fermions but it gives a logarithmically  divergent term if the fermions have a mass. This fact highlights thermodynamic incompatibility between formation of a finite chiral density and the absence of the chiral symmetry for massive fermions. In the response, the system generates an ultraviolet divergent contribution to the free energy $\Omega \sim m_0^2 \mu_5^2 \ln \Lambda/m_0$ which forces the dynamical system to vanish the chiral chemical potential $\mu_5$. In our work we suggest that this divergence may, however, be regularized and then renormalized in order to describe transient phenomena with a nonzero chiral density in the theories with dynamical mass generation (for example in quark-gluon plasma formed in heavy-ion collisions and described by QCD).
}

In order to support the need of the renormalization of the chiral chemical potential in QCD with a nearly-massless quarks, we invoked the following chain of arguments: the bare chemical potential creates a chiral charge density,
that tends to decay due to chirality flips that are catalyzed by the presence of the dynamical mass; 
the dynamical mass appears as a result of the interactions of the theory, and the interactions require the renormalization of the corresponding couplings and observables. Thus, the processes that involve the (non-)con\-ser\-vation of the chiral density are 
affected by the flow in the renormalization-group space of QCD, so that the chiral chemical potentials and the chiral charge should also be affected by the renormalization. 
This statement also applies to the theories where the quark mass appears dynamically as a result of the spontaneous breaking of chiral symmetry. 

\corr{
It will be interesting to check whether these predictions are valid also in a model with three dynamical flavors,
and to explore the behavior of different topological susceptibilities 
to probe the the axial symmetry restoration that emerges in various contexts~\cite{He:2005tf,Wang:2018gmj,Bazavov:2012qja,Kapusta:2019ktm,Suzuki:2019vzy}.
We leave these projects to near future studies.
}

\begin{acknowledgments}
The authors acknowledge Navid Abbasi, Marco Frasca and John Petrucci for inspiration, 
discussions and comments on the first version of this article.
M. R. is supported by the National Science Foundation of China (Grants No.11805087 and No. 11875153)
and by the Fundamental Research Funds for the Central Universities (grant number 862946).
\corr{M.C. is partially supported by Grant No. 0657-2020-0015 of the Ministry of Science and Higher Education of Russia.}
Z.Y.L. is supported by the Scientific Research Fund of Hunan Provincial Education Department (Grant No.~19C0772)
and by the National Science Foundation of China (Grant No.11835015). 
\end{acknowledgments}

\end{document}